\documentclass[aps,prb,fleqn,12pt,amsmath,amssymb]{revtex4}
\usepackage{epsfig}
\usepackage{color}
\usepackage{graphicx}
\usepackage{pdfpages}
\usepackage{amssymb,amsmath}
\usepackage{hyperref}

\begin{document}
\title{Spin waves and stability of zigzag order in the Hubbard model with spin-dependent hopping terms: application to the honeycomb lattice compounds {\boldmath ${\rm Na_2 Ir O_3}$} and {\boldmath ${\rm \alpha - Ru Cl_3}$}}
\author{Shubhajyoti Mohapatra}
\affiliation{Department of Physics, Indian Institute of Technology, Kanpur - 208016, India}
\author{Avinash Singh}
\email{avinas@iitk.ac.in}
\affiliation{Department of Physics, Indian Institute of Technology, Kanpur - 208016, India}
\date{\today} 

\begin{abstract}

Spin waves in the zigzag ordered state on a honeycomb lattice are investigated within a Hubbard model with spin-dependent hopping terms. Roles of the emergent Kitaev, Heisenberg, Dzyaloshinskii-Moriya, and symmetric-off-diagonal spin interactions are investigated on the stability of the zigzag order. While the DM interactions are found to destabilize the zigzag order, the secondary spin-dependent hopping terms (associated with structural distortions) are shown to strongly stabilize the zigzag order and account for magnetocrystalline anisotropy, easy axis, and spin wave gap. The calculated spin wave dispersion and energy scale are in good agreement with inelastic neutron scattering measurements on $\rm \alpha - RuCl_3$ and $\rm Na_2 Ir O_3$. 

\end{abstract}

\maketitle
\newpage

\section{Introduction}

Involving strong spin-orbit coupling (SOC), the $4d$ and $5d$ transition metal oxides are of considerable recent interest in view of potential technological importance and novel electronic and magnetic properties such as topological band or Mott insulators, quantum spin liquids, field-induced topological order, topological superconductors, and spin-orbital Mott insulators. The honeycomb lattice compounds such as $\alpha$-$\rm{RuCl_3}$ and $\rm Na_2IrO_3$, in which Ir (Ru) and O (Cl) ions form edge-sharing octahedra giving rise to O (Cl) assisted hopping between different Ir (Ru) orbitals, exhibit collinear zigzag antiferromagnetic (AFM) ground state as confirmed by both resonant magnetic X-ray scattering and neutron scattering experiments.\cite{liu_PRB_2011,choi_PRL_2012,ye_PRB_2012,johnson_PRB_2015,sears_PRB_2015}

 
Inelastic neutron scattering (INS) and resonant inelastic X-ray scattering (RIXS) experiments on the above two compounds show dispersive spin wave excitations with very low energy scale. For $\rm Na_2IrO_3$, spin wave excitations have been identified below 6 meV in INS studies with a sinusoidal decrease of energy at low $Q$,\cite{choi_PRL_2012} and found extending up to 35 meV in RIXS studies, with additional peaks in the range $0.4 < \omega < 0.8$ eV associated with the excitonic modes.\cite{gretarsson_PRB_2013} For $\alpha-{\rm RuCl_3}$, INS measurements show spin wave gap of 
$\sim 2$ meV around the ${\rm M'}$ point of the Brillouin zone.\cite{ran_PRL_2017} In both systems, the N\'{e}el temperature ($\sim 15$ K) is low compared to the Curie-Weiss temperature ($\sim -100$ K).\cite{choi_PRL_2012,sears_PRB_2015}


Band structure calculations carried out within density functional theory (DFT), dynamical mean-field theory (DMFT), and three-orbital model yield insulating AFM ground state and show significant role of SOC and electron correlation in determining the electronic structure for both compounds.\cite{foyevtsova_PRB_2013,yamaji_PRL_2014,kim_PRB_2015,kim_PRB_2016,laubach_PRB_2017} The insulating nature of $\rm Na_2IrO_3$ has been probed by angle-resolved photoemission study (ARPES),\cite{comin_PRL_2012} which reveal narrow electronic bandwidth ($\sim 100$ meV) of the filled $t_{2g}$ bands as well as large optical gap ($\sim 340$ meV), and show little variation in photoemission intensity with momentum, suggesting relatively localized electronic states. The well gapped AFM insulating state, with the gap robust even above the N\'{e}el temperature, suggests Mott (not Slater) behavior.\cite{kim_PRB_2016} In the case of $\alpha-\rm{RuCl_3}$ also, ARPES experiments show weakly dispersing Ru $d$ bands near the Fermi level and relatively large gap,\cite{zhou_PRB_2016,koitzsch_PRL_2016,sandilands_PRB_2016} establishing it as a Mott insulator. The three-orbital model parameters deduced from DFT calculations show that the largest hopping integral $t_{dd} \sim 270$ meV for ${\rm Na_2IrO_3}$ and $\sim 160$ meV for $\alpha-\rm{RuCl_3}$, which predominantly arise from hopping through the bridging ligand oxygen and halogen $p$ orbital, respectively.\cite{foyevtsova_PRB_2013,winter_PRB_2016} It should be noted that the spin-wave energy scale found experimentally is very low compared to these hopping energy scales.

Although the magnetic ions form a bipartite lattice, these systems do not exhibit the conventional N\'{e}el AFM ground state, highlighting the importance of anisotropic magnetic interactions. Arising from the spin-orbital entangled nature of the $J_{\rm eff} = 1/2$ state and orbital mixing in the electron hopping, the anisotropic magnetic interactions in these $d^5$ compounds intrinsically frustrate the N\'{e}el state and naturally allows for novel magnetic states such as zigzag, stripy, incommensurate spiral and spin liquid.\cite{chaloupka_PRL_2010} 

Most of the earlier studies of magnetic ordering, excitations, and anisotropy in these systems have been carried out in terms of low energy effective spin models with various anisotropic spin interactions. The celebrated Kitaev-Heisenberg (KH) model, with the unusual bond dependent Kitaev exchange as the dominant interaction, has been studied extensively.\cite{chaloupka_PRL_2010,chaloupka_PRL_2013,plumb_PRB_2014,sandilands_PRL_2015,kim_PRB_2015,chaloupka_PRB_2016,ysingh_PRL_2012,rau_PRL_2014,chaloupka_PRB_2015,ran_PRL_2017,wang_PRB_17,janssen_arxiv_2017}
Within the nearest-neighbor Kitaev-Heisenberg (nnKH) model, a zigzag ordered ground state is obtaned  for antiferro Kitaev interaction ($K > 0, J < 0$), whereas ferro Kitaev interaction ($K < 0$) yields stripy order provided $J > 0$ and $|K|>|J|$. Model calculations focusing on a selection of these interactions such as $K,J,\Gamma$ have predicted phase diagrams hosting a rich variety of classical broken-symmetry states.\cite{chaloupka_PRL_2013,rau_PRL_2014,sizyuk_PRB_2014,katukuri_NJP_2014,chaloupka_PRB_2015} In each case, the specific ground state is selected by a competition between the various interactions, so that none of the terms can be neglected a priori. 

It was subsequently suggested that the experimentally observed zigzag order could be realized in the nnHK model with ferro Kitaev interaction when supplemented with a sizable spin-off-diagonal (SOD) interaction $\Gamma$.\cite{rau_PRL_2014,chaloupka_PRB_2015,ran_PRL_2017,wang_PRB_17,janssen_arxiv_2017} 
Ferro sign of Kitaev interaction was motivated by perturbative expansion studies around various strong coupling limits within the t$_{2g}$ sector in terms of electron hopping integrals, on-site Coulomb interaction $U$, and Hund’s coupling $J_H$.\cite{rau_PRL_2014,sizyuk_PRB_2014,yamaji_PRL_2014,winter_PRB_2016} However, a clear picture of the appropriate spin interactions for the two relevant compounds $\rm Na_2 Ir O_3$ and $\rm Ru Cl_3$ has still not emerged. 

Magnetic properties of the honeycomb lattice $d^5$ compounds have not been theoretically discussed in terms of an effective itinerant electron model with spin-dependent hopping terms. The single-band correlated electron model within the basis of the spin-orbital entangled states of the $J=1/2$ sector provides a more compact description of the SOC induced magnetic anisotropy, as seen from the several anisotropic spin interaction terms (Kitaev, Dzyaloshinski-Moriya, symmetric off-diagonal) generated within the strong coupling expansion. Although the spin-dependent hopping terms can arise in general from the orbital mixing between the $t_{2g}$ orbitals due to the O-assisted (or Cl-assisted) hopping, we will treat these as effectively incorporating the anisotropic magnetic interactions.

In this paper, we will therefore investigate spin dynamics in the zigzag state within a Hubbard model on the honeycomb lattice with spin-dependent hopping terms. This will allow for a microscopic and unified understanding of the roles of the emergent Kitaev, Heisenberg, Dzyaloshinskii-Moriya (DM), and symmetric-off-diagonal (SOD) interactions on the stabilization of zigzag order and measured spin wave excitations in the honeycomb lattice compounds. Spin wave dispersion in these compounds have been investigated within the three-orbital Hubbard model,\cite{igarashi_JESRP_2016} but not within the Kitaev-Hubbard model.\cite{liang_PRB_2014}  

\begin{figure}
\vspace*{0mm}
\hspace*{0mm}
\psfig{figure=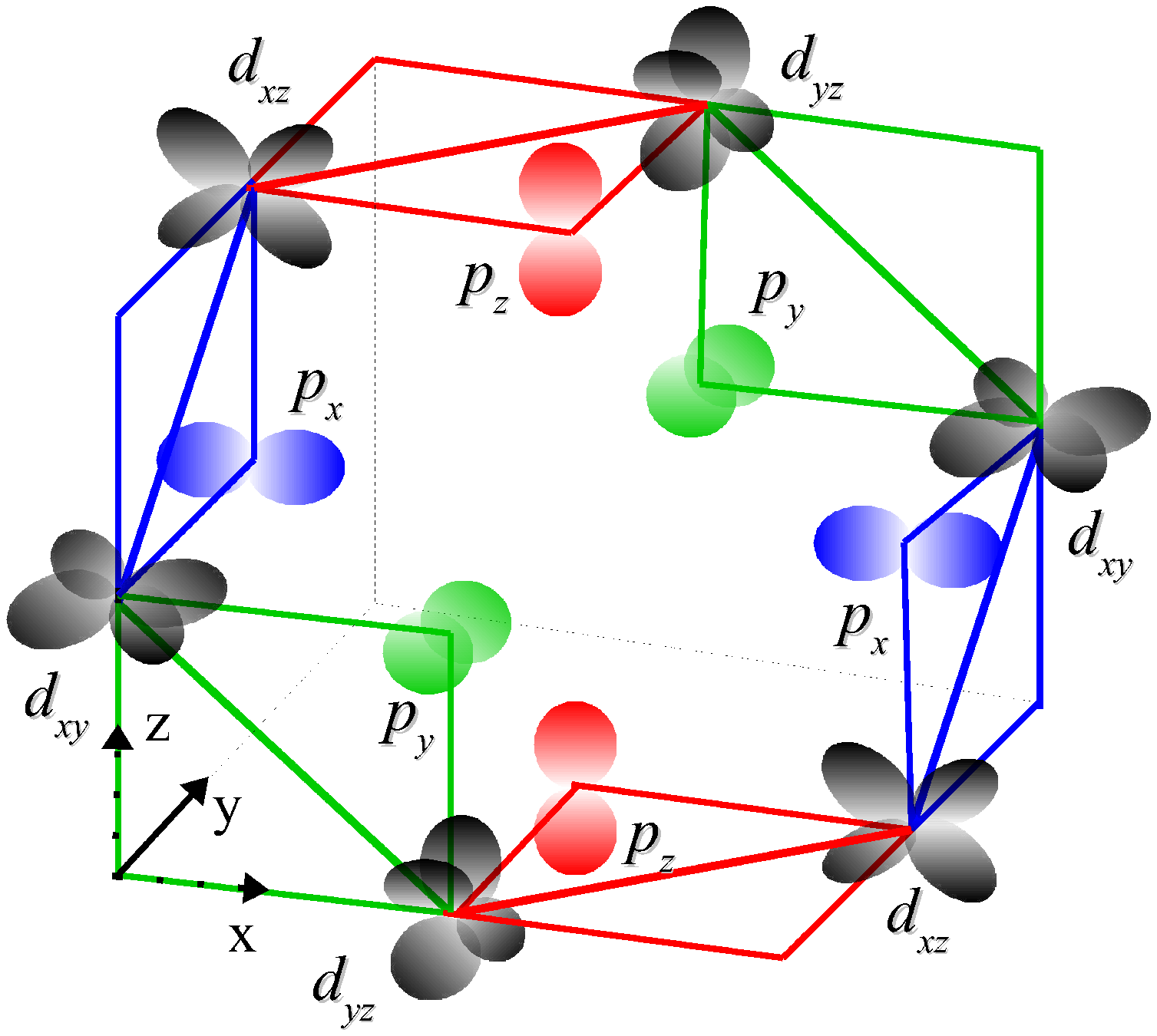,angle=0,width=70mm} \hspace*{10mm}
\psfig{figure=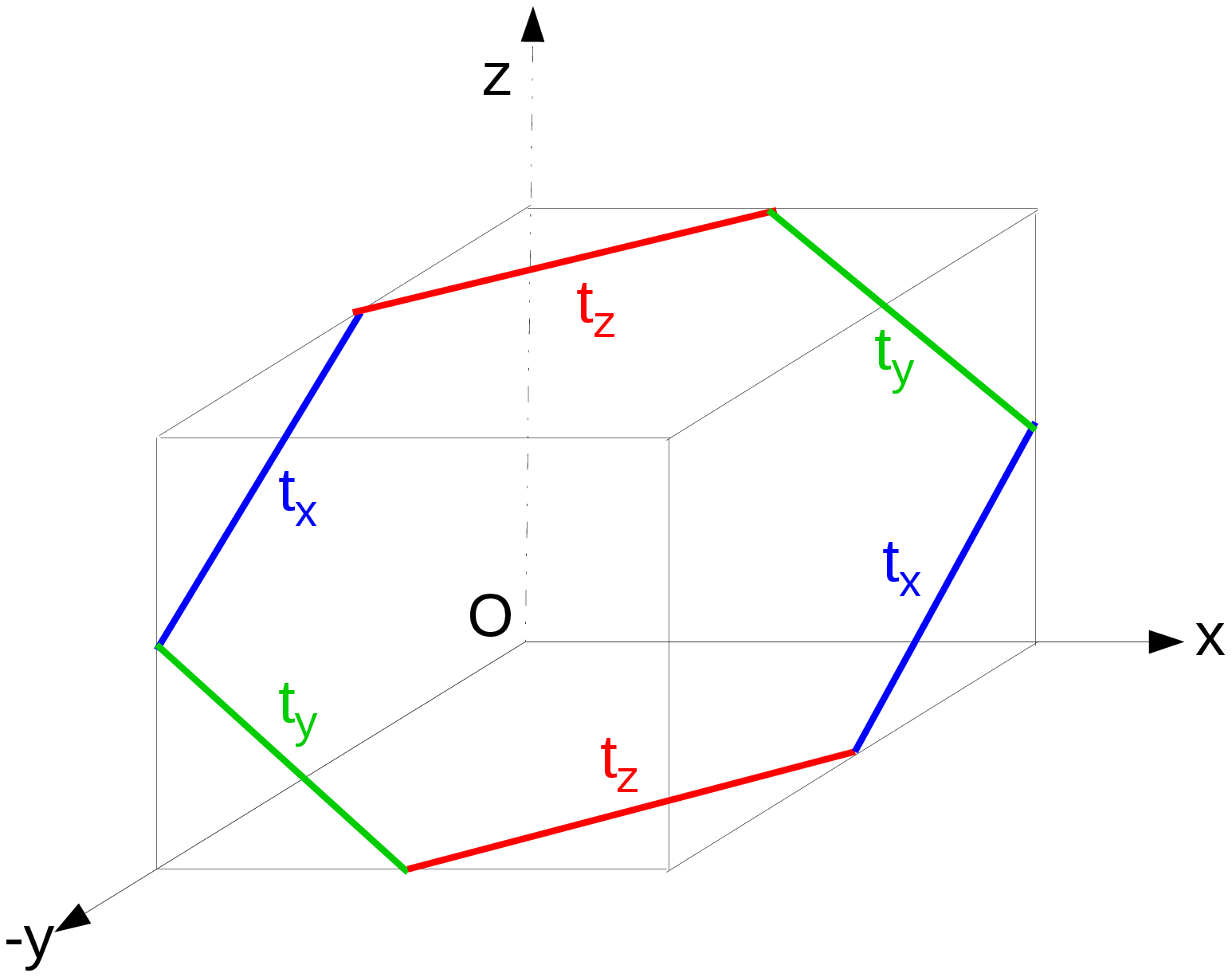,angle=0,width=60mm} 
\caption{Schematic representation of the honeycomb lattice structure of Ir in the cubic setting. For each of the three types of Ir-Ir bonds, O-assisted hopping between only two particular $t_{\rm 2g}$ orbitals is possible: $d_{yz} - d_{xz}$ (red), $d_{xz} - d_{xy}$ (blue), and $d_{xy} - d_{yz}$ (green). Two possible hopping paths via oxygen $p$ orbitals are shown. Similar Cl-assisted hopping leads to orbital mixing in $\rm \alpha - Ru Cl_3$.} 
\label{hc_embedding}
\end{figure}

\section{Orbital mixing and spin-dependent hopping}

Honeycomb lattice iridates contain edge-sharing $\rm IrO_6$ octahedra, and relative orientations of neighboring $\rm IrO_6$ octahedra give rise to different dominant hopping pathways between the $t_{\rm 2g}$ orbitals located on Ir ions at the centre of these octahedra.\cite{jackeli_PRL_2009} There are two Ir-O-Ir hopping paths with 90$^\circ$ bonding geometry between two neighboring Ir ions (in the ideal cubic setting), and the O-assisted hopping ($\pi$ overlap) between unlike Ir orbitals gives rise to effective orbital mixings. Specifically, dominant orbital mixing hopping terms between $d_{xz}-d_{yz}$, $d_{yz}-d_{xy}$, $d_{xy}-d_{xz}$ orbitals are induced by the O-assisted hopping for nearest-neighbor Ir sites in the same $z$, $y$, $x$ planes, respectively (Figure \ref{hc_embedding}).

Transforming from the three-orbital basis to the $J_{\rm eff}$ basis, where the 3/2 and 1/2 sectors are well separated in the strong SOC limit, the orbital mixing hopping terms result in spin-dependent hopping terms $t_z, t_y, t_x$ in the $J_{\rm eff}=1/2$ sector.\cite{iridate_paper} In the ideal cubic setting, the two orbital-mixing hopping terms between neighboring Ir ions via the two O-assisted hopping pathways exactly cancel. The net small $t_\mu$ is therefore ascribed to avoided cancellation resulting from structural distortions, as discussed later.


Spin-dependent hopping terms have been suggested as possible source of anisotropic interactions (such as the first neighbor Kitaev exchange) in the honeycomb lattice compound $\rm Na_2IrO_3$,\cite{foyevtsova_PRB_2013} and were considered earlier to study topologically nontrivial electronic states,\cite{shitade_PRL_2009} where they were ascribed to the asymmetry between two hopping paths connecting a second neighbor Ir pair. Considering effective spin interactions (Appendix A) corresponding to second neighbor spin-dependent hopping terms, the zigzag, stripy, and N\'{e}el states are degenerate at the classical level. Therefore, in the following, we will consider for simplicity only first neighbor spin-dependent hopping terms which strongly favor the zigzag state. 

We will therefore consider the one-band Hubbard model on the honeycomb lattice:
\begin{equation}
H = -t_1 \sum_{\langle ij \rangle,\sigma} (a_{i\sigma} ^\dagger a_{j\sigma} + H.c.)
-t_2 \sum_{\langle \langle ij \rangle \rangle,\sigma} (a_{i\sigma} ^\dagger a_{j\sigma} + H.c.)
- \sum_{\langle ij \rangle,\mu} \psi_i ^\dagger [i \sigma_\mu t_\mu] \psi_j  
+ U \sum_i a_{i\uparrow} ^\dagger a_{i\uparrow} a_{i\downarrow} ^\dagger a_{i\downarrow}
\label{sp_dep_Hubbard}
\end{equation}
where $t_1,t_2$ are the first and second neighbor usual hopping terms, $t_\mu$ represent the spin-dependent hopping terms which are alternately $t_x$ / $t_y$ / $t_z$ around each honeycomb plaquette as shown in Fig. \ref{zigzag}, and $U$ is the local Coulomb interaction term. The spin-dependent hopping terms $t_\mu$ are restricted to only first neighbors for simplicity. Here $a_{i\sigma}$ represents the electron annihilation operator and $\psi_i = (a_{i\uparrow} \; a_{i\downarrow})$ for site $i$. In the following, we will set $t_\mu = 1$ as the overall energy scale. 

\section{zigzag state on the honeycomb lattice}

\begin{figure}
\vspace*{-20mm}
\hspace*{0mm}
\psfig{figure=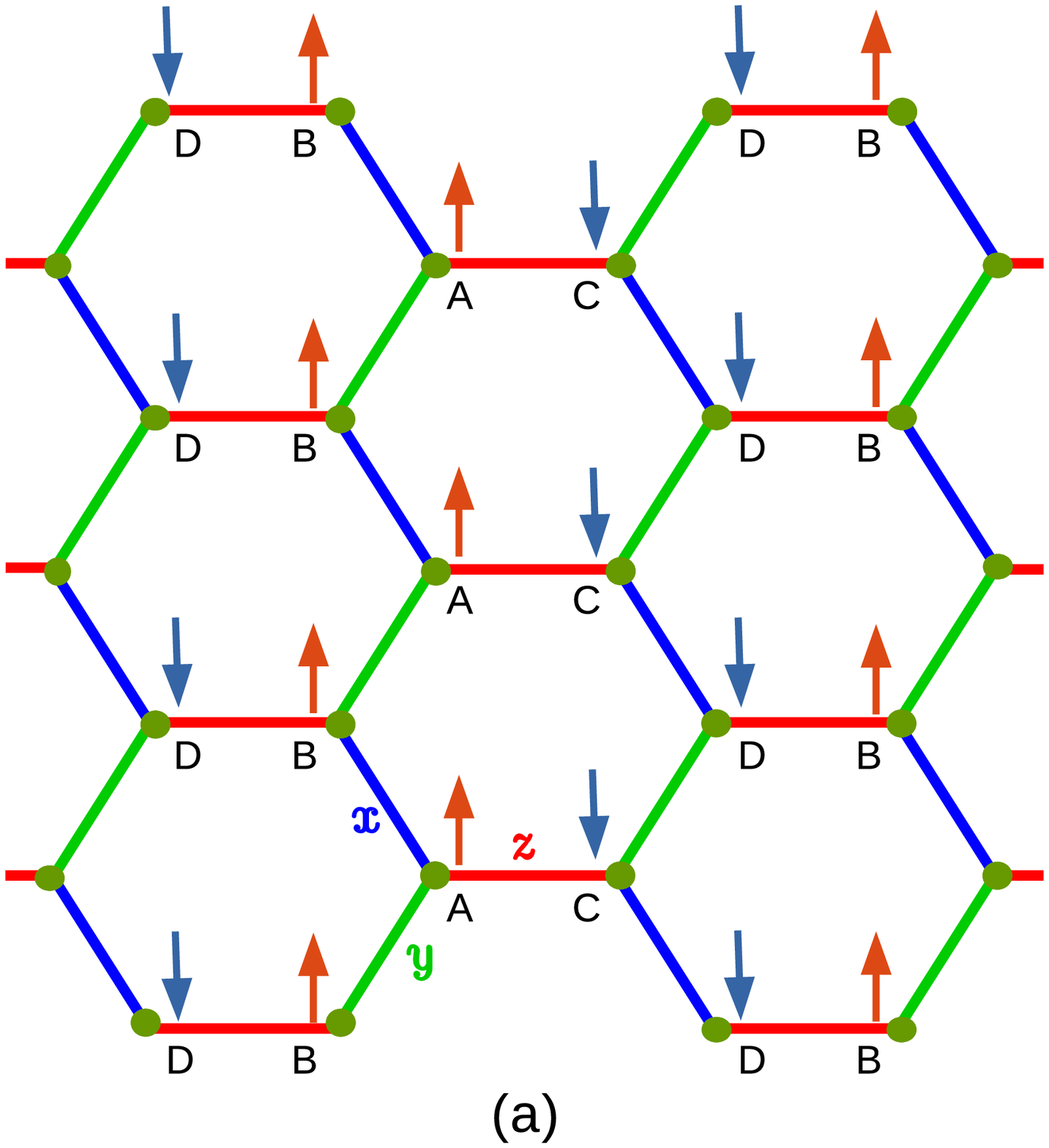,angle=0,width=52mm} 
\psfig{figure=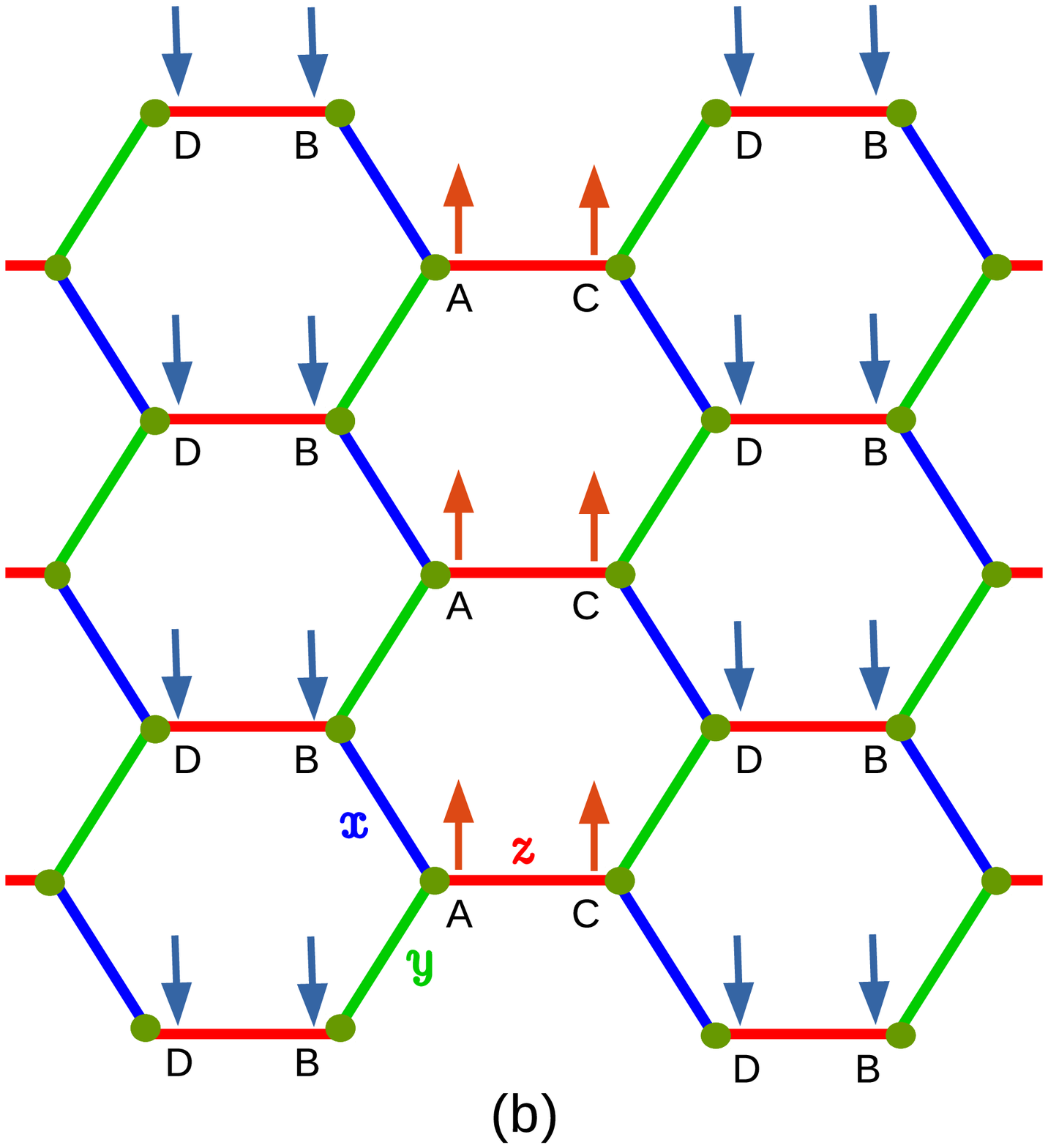,angle=0,width=52mm} 
\psfig{figure=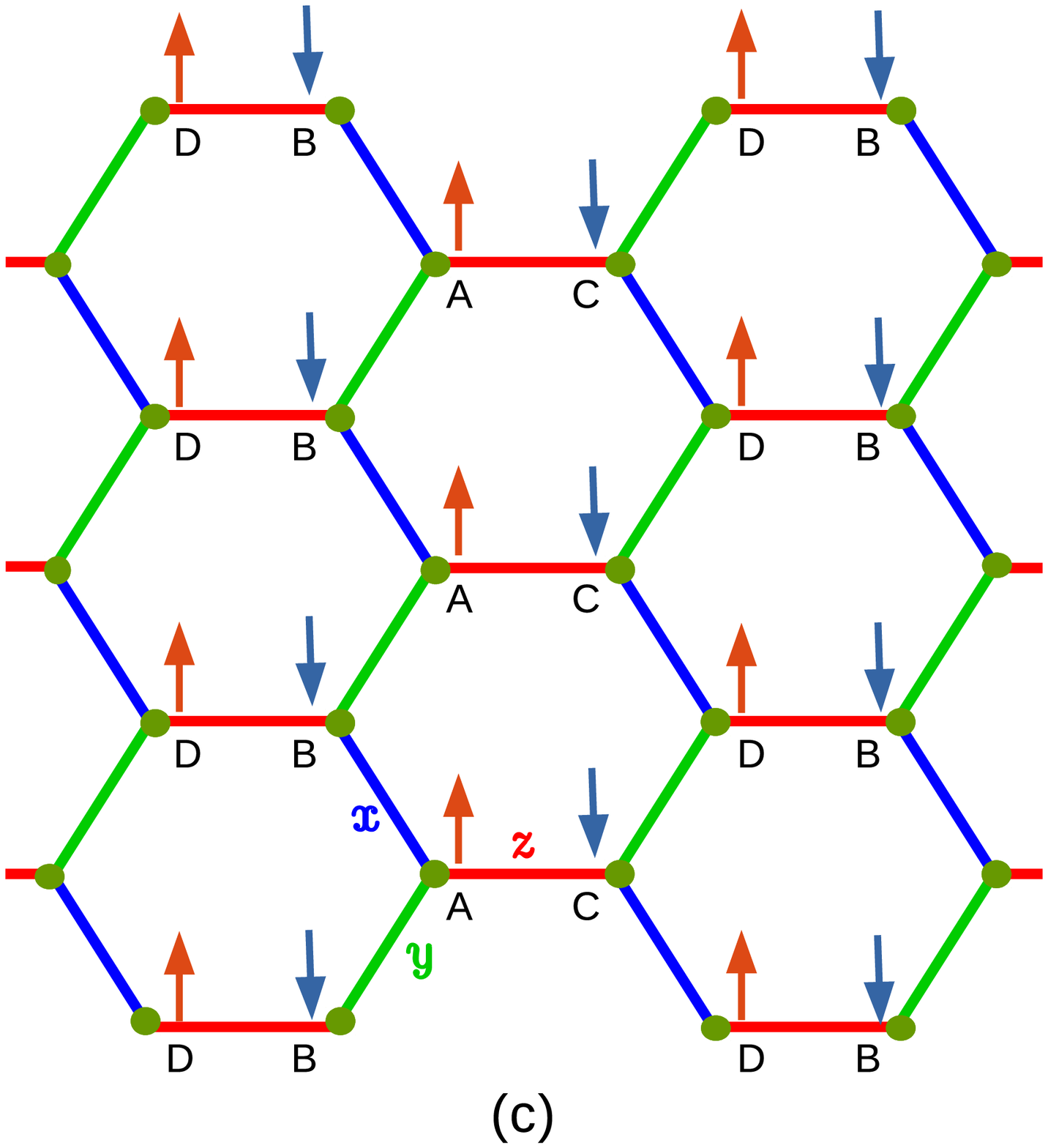,angle=0,width=52mm} 
\vspace{-15mm}
\caption{The zigzag (a), stripy (b), and N\'{e}el (c) magnetic orders on the honeycomb lattice, showing the four sublattices (A,B,C,D). The spin-dependent hopping terms are alternately $t_x$ / $t_y$ / $t_z$ around each honeycomb plaquette, as indicated by $x,y,z$ with blue, green, and red bonds.} 
\label{zigzag}
\end{figure}

We now consider the above Hamiltonian (\ref{sp_dep_Hubbard}) in the Hartree-Fock (HF) approximation corresponding to magnetic ordering in the $z$ direction. In a composite four-sublattice, two-spin basis corresponding to the zigzag magnetic order shown in Fig. \ref{zigzag}(a), the HF Hamiltonian matrix is obtained as:

{\small
\begin{equation}
H_{\rm HF} ^\sigma (\bf k) = \left [ 
\begin{array}{cccccccc}
\varepsilon_{k2}^{(b)} - \Delta & 0 & \varepsilon_{k1}^{(a)} &\varepsilon_{kx} + \varepsilon_{ky} & \varepsilon_{k1}^{(b)} + \varepsilon_{kz} & 0 & \varepsilon_{k2}^{(a)} & 0 \\
  & \varepsilon_{k2}^{(b)} + \Delta & \varepsilon_{kx} - \varepsilon_{ky} & \varepsilon_{k1}^{(a)} & 0 & \varepsilon_{k1}^{(b)} - \varepsilon_{kz} & 0 & \varepsilon_{k2}^{(a)} \\
  &  & \varepsilon_{k2}^{(b)} - \Delta & 0 & \varepsilon_{k2}^{(a)} & 0 & \varepsilon_{k1}^{(b)*} + \varepsilon_{kz}^* & 0 \\
  &  &  & \varepsilon_{k2}^{(b)} + \Delta & 0 & \varepsilon_{k2}^{(a)} & 0 & \varepsilon_{k1}^{(b)*} - \varepsilon_{kz}^* \\
  &  &  &  & \varepsilon_{k2}^{(b)} + \Delta & 0 & \varepsilon_{k1}^{(a)*} & \varepsilon_{kx}^* - \varepsilon_{ky}^*  \\ 
  &  &  &  &  & \varepsilon_{k2}^{(b)} - \Delta & \varepsilon_{kx}^* + \varepsilon_{ky}^* & \varepsilon_{k1}^{(a)*} \\ 
  &  &  &  &  &  & \varepsilon_{k2}^{(b)} + \Delta & 0 \\ 
  &  &  &  &  &  &  & \varepsilon_{k2}^{(b)} - \Delta \\ 
\end{array} \right ]
\label{hf_hamiltonian}
\end{equation} 
}
where the band terms corresponding to normal and spin-dependent hoppings in different directions are given by:
\begin{eqnarray}
\varepsilon_{k1}^{(a)} & = & -2t_1 e^{-i k_x/2} \cos (\sqrt{3}k_y/2) \nonumber \\
\varepsilon_{k1}^{(b)} & = & -t_1 e^{i k_x} \nonumber \\
\varepsilon_{k2}^{(a)} & = & -4t_2 \cos (3 k_x/2) \cos (\sqrt{3} k_y/2) \nonumber \\
\varepsilon_{k2}^{(b)} & = & -2t_2 \cos (\sqrt{3}k_y) \nonumber \\
\varepsilon_{kx} & = & -it_x e^{-i (k_x /2 - \sqrt{3} k_y/2)} \nonumber \\
\varepsilon_{ky} & = & -t_y e^{-i (k_x /2 + \sqrt{3} k_y/2)} \nonumber \\
\varepsilon_{kz} & = & -it_z e^{i k_x}
\end{eqnarray}
and $\Delta = mU/2$ is the staggered field in terms of the sublattice magnetization:
\begin{equation}
m (\Delta) 
= (n_\uparrow ^A - n_\downarrow ^A)(\Delta) 
= (n_\uparrow ^B - n_\downarrow ^B)(\Delta) 
= (n_\downarrow ^C - n_\uparrow ^C)(\Delta)
= (n_\downarrow ^D - n_\uparrow ^D)(\Delta)
\end{equation}
which is determined self-consistently from the electronic densities $n_\sigma$ calculated from $H_{\rm HF} ^\sigma (\bf k)$ for the two spins $\sigma$=$\uparrow,\downarrow$ on the four magnetic sublattices. In practice, it is easier to choose $\Delta$ and determine $U$ from the calculated sublattice magnetization $m(\Delta)$. In the large $U$ limit, $2\Delta \approx U$ as $m \rightarrow 1$. We will consider only the half-filled case ($n=1$) with Fermi energy in the AFM band gap. 
 
\section{Spin waves in the zigzag AFM state}

In the following, we will consider transverse spin fluctuations in the $x$ and $y$ directions in the zigzag ordered state (Fig. \ref{zigzag}) with magnetic ordering in the $z$ direction. These low-energy collective excitations, studied earlier within the Kitaev-Heisenberg localized spin models,\cite{chaloupka_PRL_2010} will include the gapless Goldstone mode corresponding to specific spin twisting modes which transform the ground state from one zigzag domain to another (Appendix B), and will also show the spin wave gap resulting from the intrinsic magnetic anisotropy due to the spin-dependent hopping terms (Appendix A).    

Due to the spin mixing terms in Hamiltonian (\ref{sp_dep_Hubbard}), spin is not a good quantum number, and we therefore use the method developed earlier for the 120$^\circ$ (non-collinear) ordering on the triangular lattice to investigate transverse spin fluctuations.\cite{tri} We consider the time-ordered transverse spin fluctuation propagator: 
\begin{equation}
[\chi ({\bf q},\omega)]^{\mu\nu} _{\alpha \beta} =
i \int dt  \; e^{i\omega (t-t')} \sum_j e^{-i{\bf q}.({\bf r}_i - {\bf r}_j)} 
\langle \Psi_{\rm G} |{\rm T}[S_i ^\mu (t) S_j ^\nu (t')]\Psi_{\rm G} \rangle
\end{equation}
where $S_i ^\mu = (1/2)\psi_i ^\dagger \sigma^\mu \psi_i$ is the spin operator, $\mu,\nu=x,y$ are the transverse spin directions, and $\alpha,\beta$ are the sublattice indices corresponding to sites $i,j$. In the random phase approximation (RPA),\cite{tri}  
\begin{equation}
[\chi ({\bf q},\omega)]  = \frac{[\chi^0 ({\bf q},\omega)]}
{{\bf 1} - 2U[\chi^0 ({\bf q},\omega)]}
\label{chi_rpa}
\end{equation}
where the bare particle-hole propagator $[\chi^0 ({\bf q},\omega)]$ is evaluated by integrating out the fermions in the broken-symmetry state. For the half-filled insulating state, the added hole $(-)$ / particle $(+)$ states lie in the lower / upper Hubbard band, and we obtain:
\begin{equation}
[\chi^0({\bf q},\omega)]^{\mu\nu} _{\alpha \beta} = \frac{1}{4}
\sum_{{\bf k}lm} \frac{\langle \sigma^\mu \rangle_\alpha ^{-+} \langle \sigma^\nu \rangle_\beta ^{-+*}}
{E_{{\bf k-q},m}^+ - E_{{\bf k},l}^- + \omega} 
+ 
\frac{\langle \sigma^\mu \rangle_\alpha ^{+-} \langle \sigma^\nu \rangle_\beta ^{+-*}}
{E_{{\bf k},l}^+ - E_{{\bf k-q},m}^- - \omega} \; ,
\end{equation}
where $E_{{\bf k},l}$ are the eigenvalues of the HF Hamiltonian matrix (\ref{hf_hamiltonian}) for  branch index $l$ and $\langle \sigma^\mu \rangle_\alpha$ denotes the spin matrix element for component $\mu = x,y$ on the $\alpha$ sublattice between particle-hole states:
\begin{equation}
\langle \sigma^\mu \rangle_\alpha ^{-+} \equiv
\langle ({\bf k-q},m)^+ |\sigma^\mu|({\bf k},l)^-\rangle_\alpha.
\end{equation}

The bare particle-hole propagator $[\chi^0({\bf q},\omega)]^{\mu\nu} _{\alpha,\beta}$ is a $[8 \times 8]$ Hermitian matrix in the composite four-sublattice ($\alpha$=A,B,C,D) and two-component ($\mu=x,y$) basis. The (real) eigenvalues $\lambda_{\bf q}(\omega)$ and eigenvectors $|\phi_{\bf q}(\omega)\rangle$  contain information about the collective spin wave modes. Also included are the particle-hole (Stoner) excitations across the band gap, given by the poles of $[\chi^0({\bf q},\omega)]^{\mu\nu} _{\alpha,\beta}$. In the following, we will focus on the spin-wave energies $\omega_{\bf q}$, which are obtained from the poles $1-2U\lambda_{\bf q}(\omega_{\bf q}) = 0$ of Eq. (\ref{chi_rpa}).

It is instructive to start with the strong-coupling expansion and consider the various spin interaction terms generated up to second order in the hopping terms (Appendix A). The DM term is absent for $t=0$, resulting in the Kitaev-Heisenberg model, for which there are three equivalent zigzag domains, with specific spin twisting modes transforming from one domain to another (Appendix B). 


\begin{figure}[htbp]
\hspace*{-0mm}
\begin{minipage}[b]{0.45\linewidth}
\centering
\psfig{figure=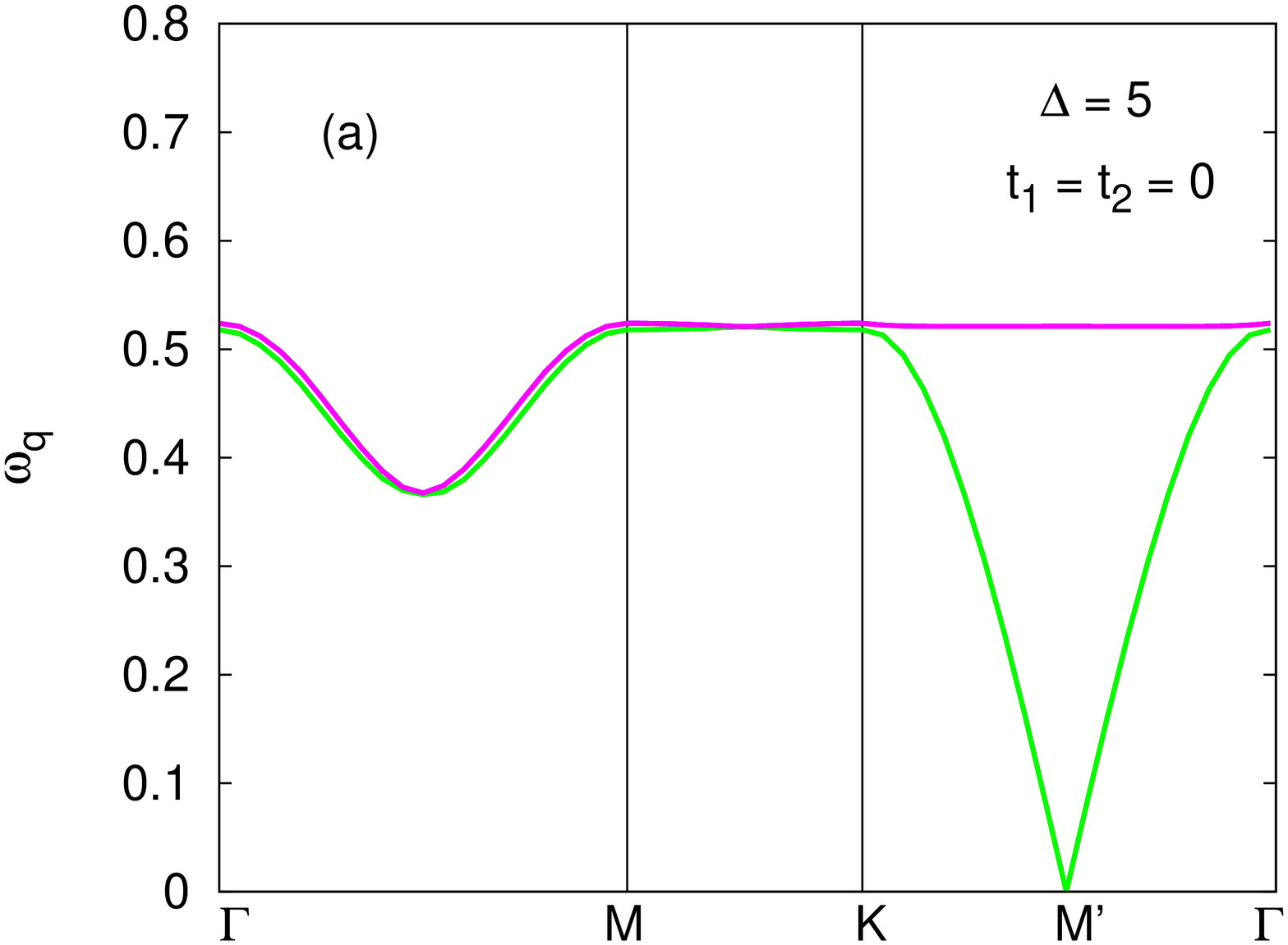,angle=-90,width=75mm}
\end{minipage}
\hspace{0.8cm}
\begin{minipage}[b]{0.45\linewidth}
\centering
\psfig{figure=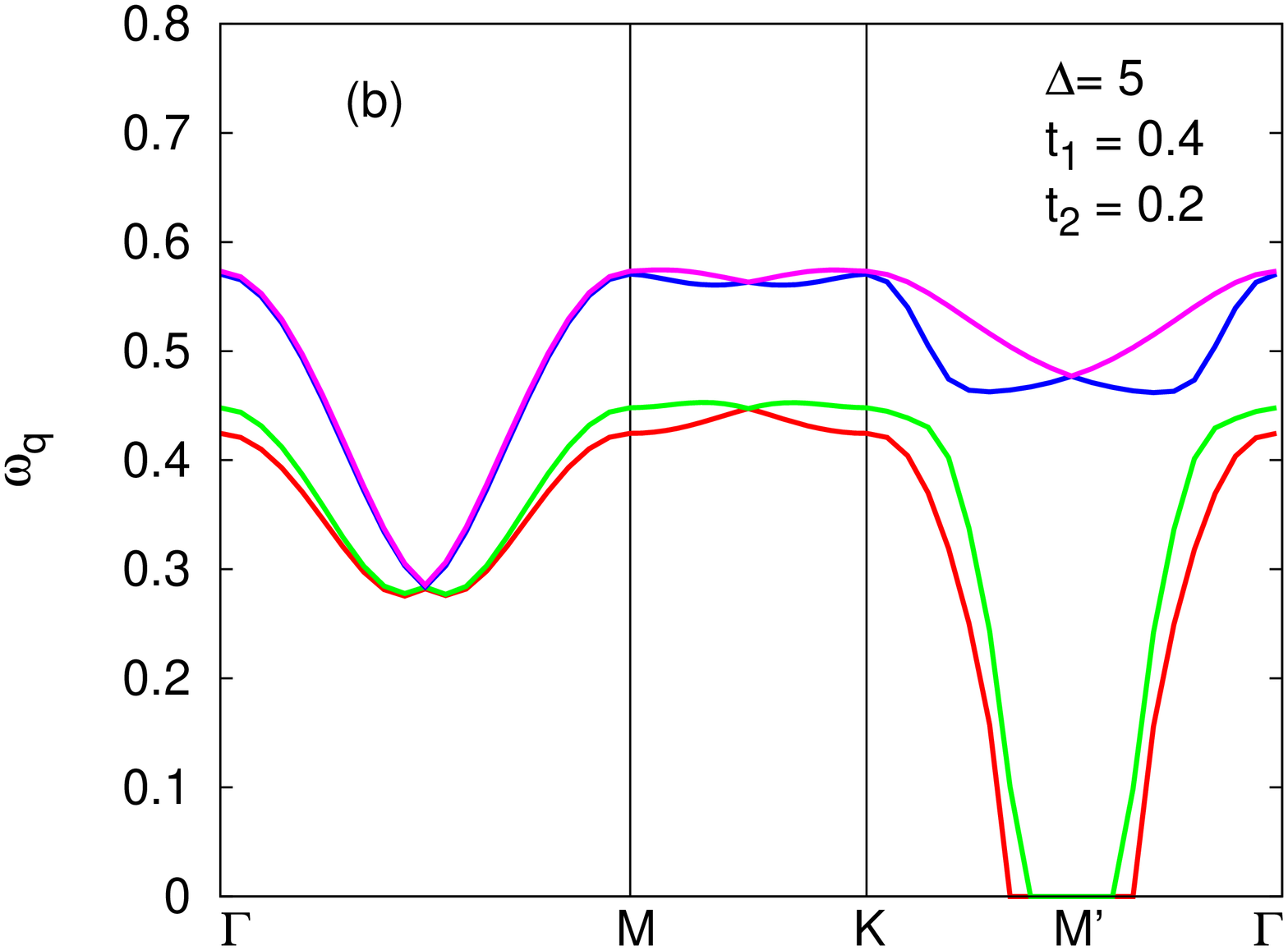,angle=-90,width=75mm}
\end{minipage} \\
\hspace*{-0mm}
\begin{minipage}[b]{0.45\linewidth}
\centering
\psfig{figure=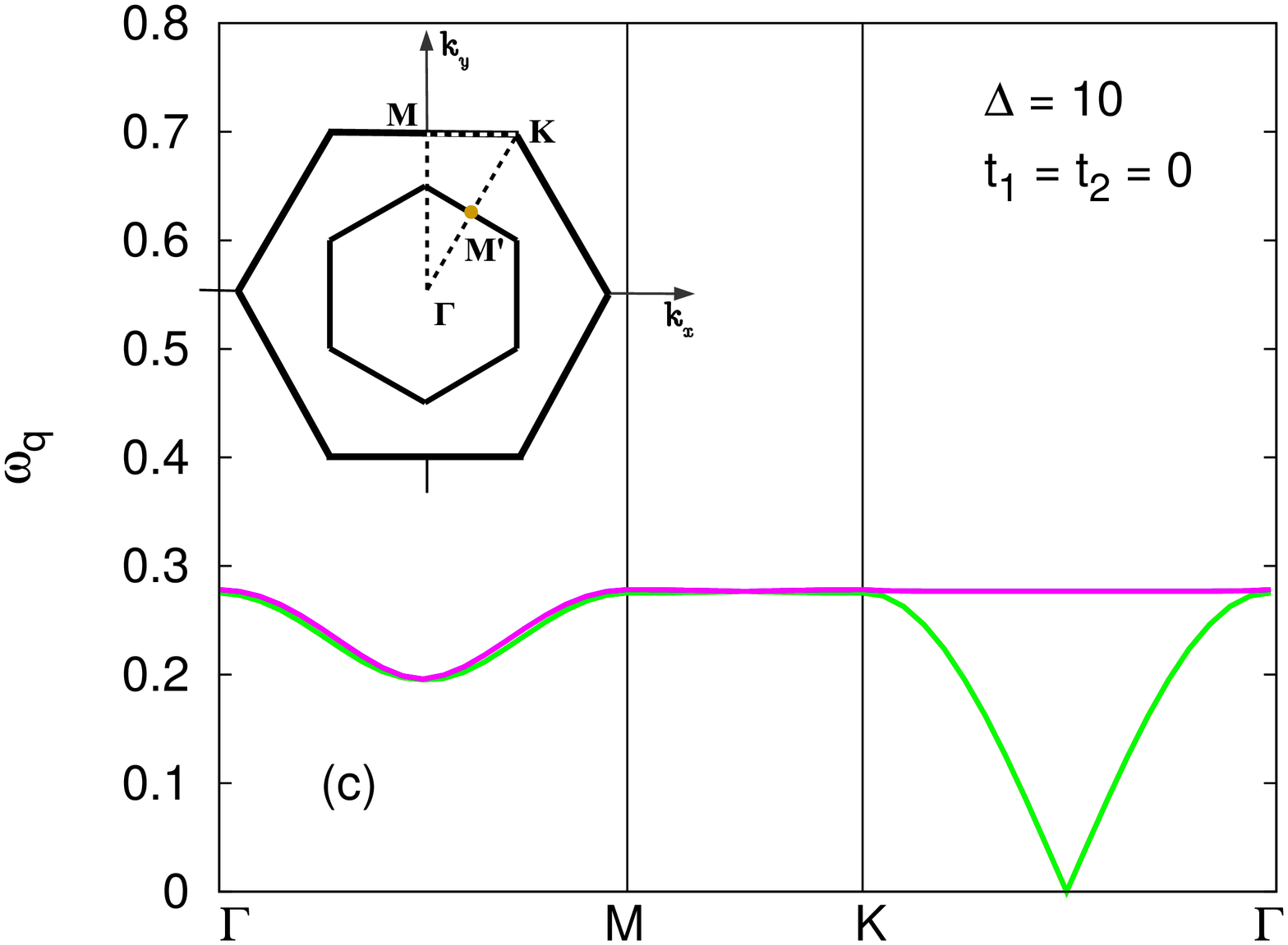,angle=0,width=67mm} \vspace*{-52mm}
\end{minipage}
\hspace{0.8cm}
\begin{minipage}[b]{0.45\linewidth}
\centering
\psfig{figure=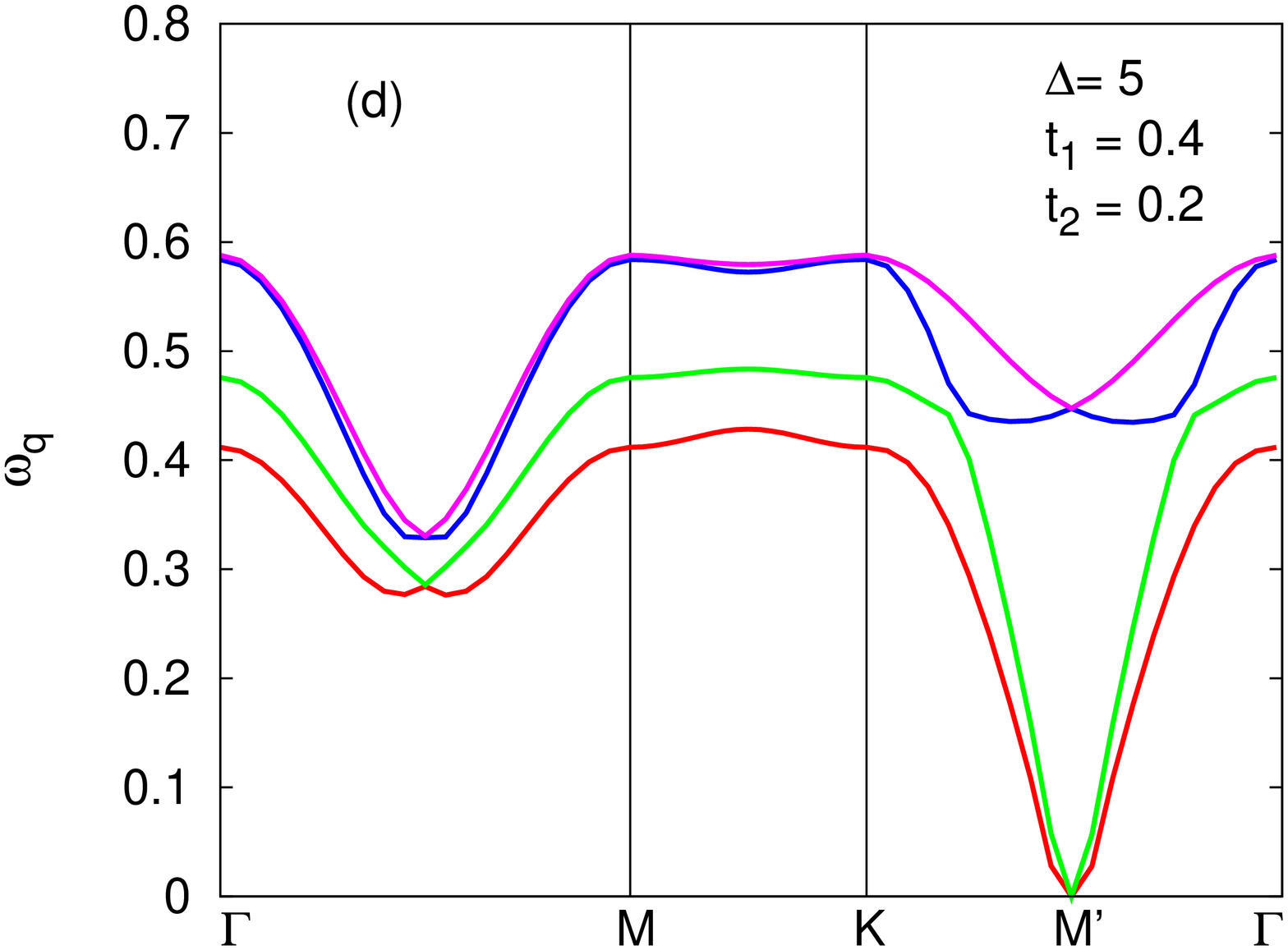,angle=-90,width=75mm} 
\end{minipage} \vspace*{5mm}
\caption{Calculated spin wave energies in the zigzag ordered state with (a,c) only spin-dependent hopping terms $t_\mu = 1$, showing the Goldstone mode and approximate scaling as $t^2/U$. Instability on including finite $t_1$ and $t_2$ due to the DM term (b) is removed when the hopping term $i\sigma_\mu t_\mu$ is replaced (d) by $\sigma_\mu t_\mu$. Inset in (c) shows the extended and magnetic Brillouin zones.}
\label{fig3}
\end{figure}

The calculated spin-wave dispersion $\omega_{\bf q}$ along different symmetry directions of the Brillouin zone are discussed below. Figs. \ref{fig3} (a,c) show the zero-energy mode at the ${\rm M'}$ point (midway between K and $\Gamma$) for the case $t_1=t_2=0$ (no DM interaction and no magnetic frustration). This Goldstone mode corresponds to the spin twisting mode (Appendix B), which costs zero energy even in the presence of magnetic anisotropy. The overall softening of the spin wave spectrum (lower branch) when $t$ is turned on reflects the frustration effect due to the Heisenberg interaction $(4t ^2/U){\bf S_i}.{\bf S_j}$ [Figs. \ref{fig3} (b,d)]. In Fig. \ref{fig3} (b), the strong instability near ${\rm M'}$, indicated by spin wave energy turning negative, is exclusively due to the DM interactions (which require both $t$ and $t_\mu$ to be non-zero). This is consistent with the absence of this instability for $t_1 = t_2 = 0$. Furthermore, for $t_2=0$, we find that this instability develops {\em for any finite} $t_1$, which cannot be accounted in terms of the Kitaev and Heisenberg interactions only (Appendix B). As DM interactions are intrinsically present in our Hubbard model analysis, a stronger destabilization of zigzag order is obtained than in earlier works on the KH model.

There is no instability at ${\rm M'}$ [Fig. \ref{fig3} (d)] if the spin-dependent hopping term $i\sigma_\mu t_\mu$ is replaced by $\sigma_\mu t_\mu$ in Eq. (\ref{sp_dep_Hubbard}). The effective spin model in this case is the KH model without the DM interaction, and the dramatic change from Fig. \ref{fig3} (b) to (d) explicitly confirms the role of DM interaction in the destabilization of zigzag order. Such spin-dependent hopping term $(\sigma_\mu t_\mu)$ was originally proposed for realizing the Kitaev spin model in cold atom systems and studied for cubic and honeycomb lattices,\cite{duan_PRL_2003,hassan_PRL_2013,rachel_PRL_2015} and has also been studied recently for the rare earth materials such as $\rm YbMgGaO_4$ and $\rm Ba_3IrTi_2O_9$, both of which are triangular lattice systems with strong SOC and possibly host spin liquid ground state.\cite{li_PRB_2016} 
 
\subsection{Stabilization of zigzag order}

\begin{figure}
\vspace*{0mm}
\hspace*{0mm}
\psfig{figure=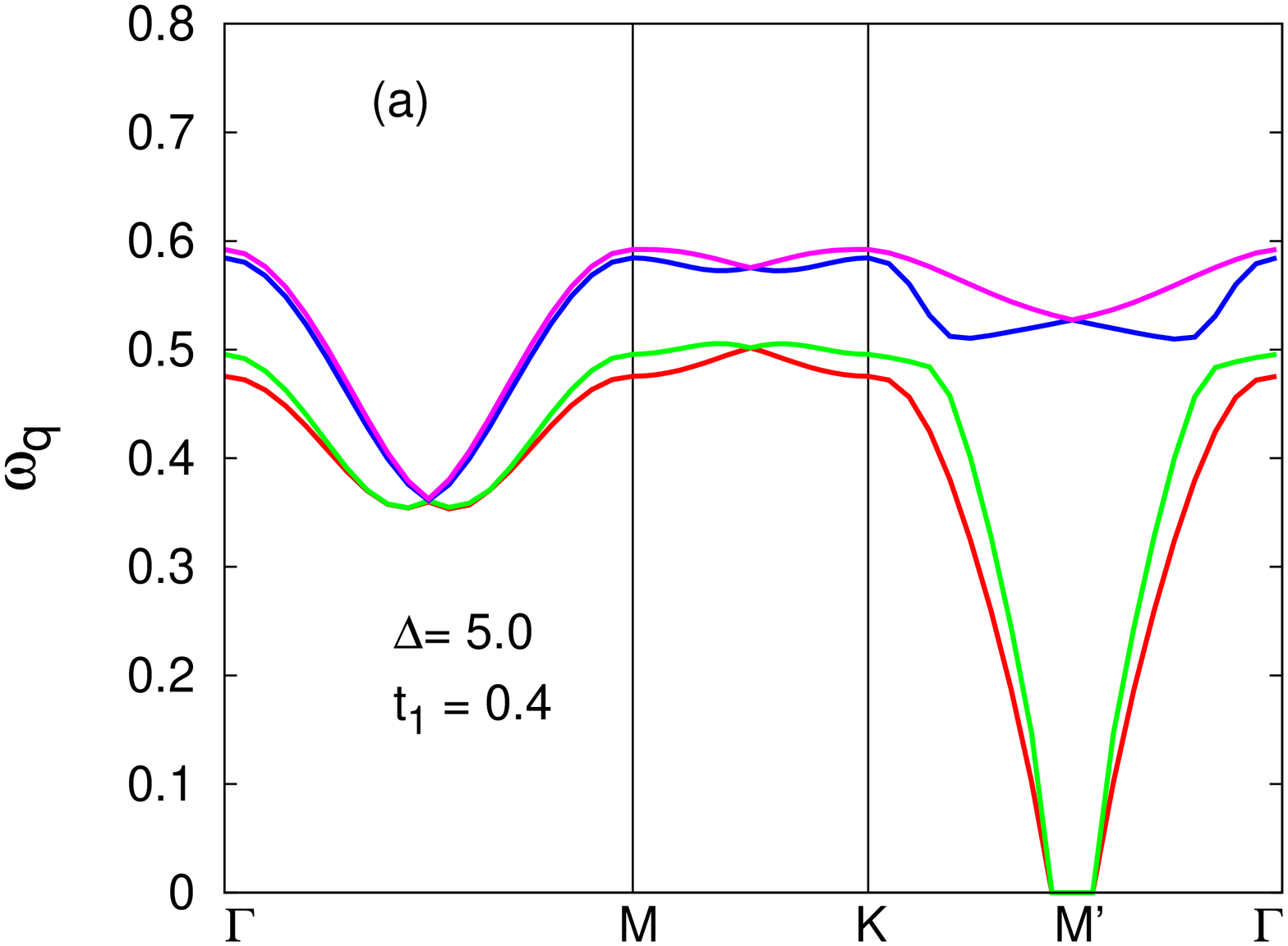,angle=-90,width=50mm} 
\psfig{figure=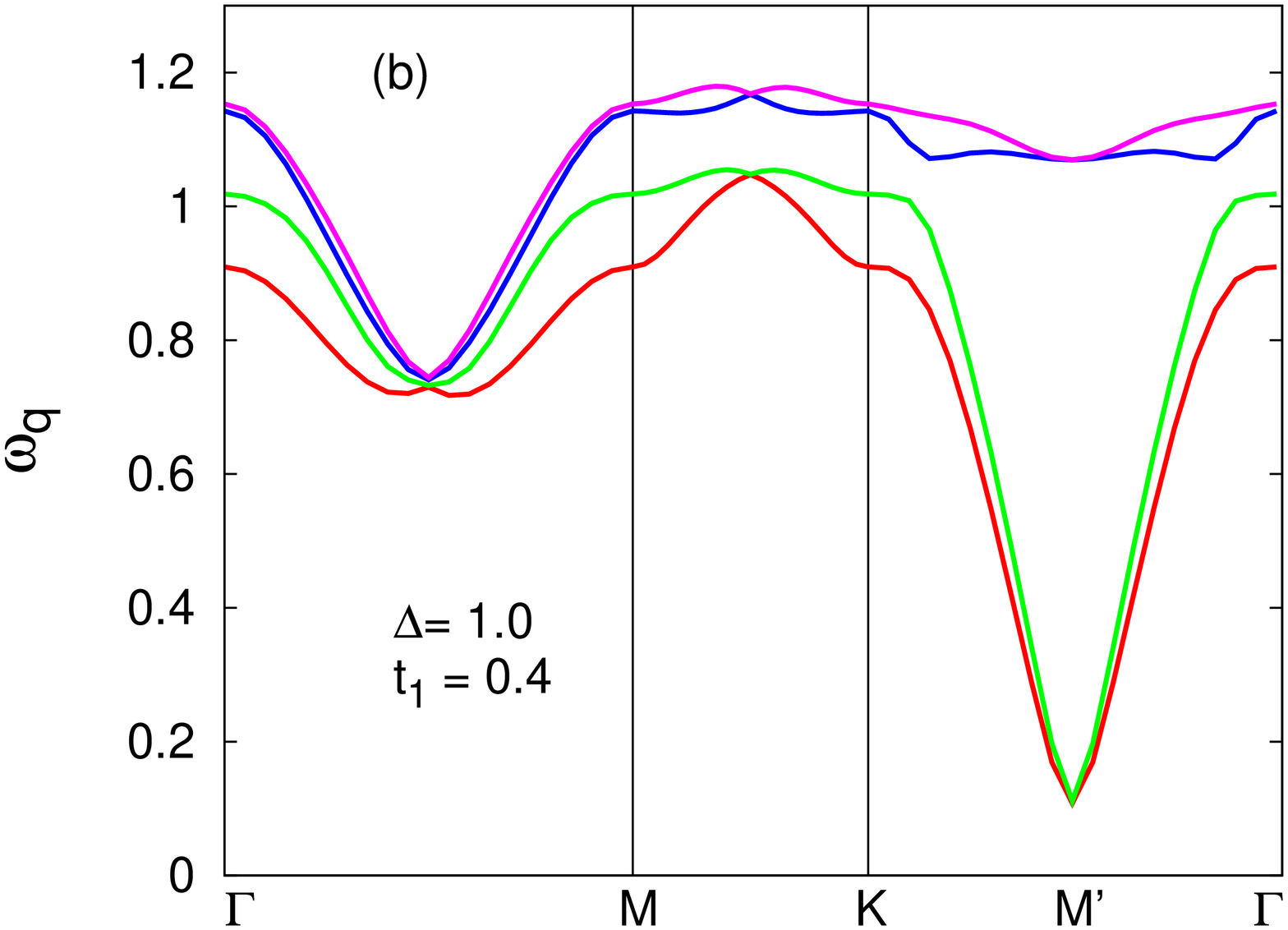,angle=-90,width=50mm} 
\psfig{figure=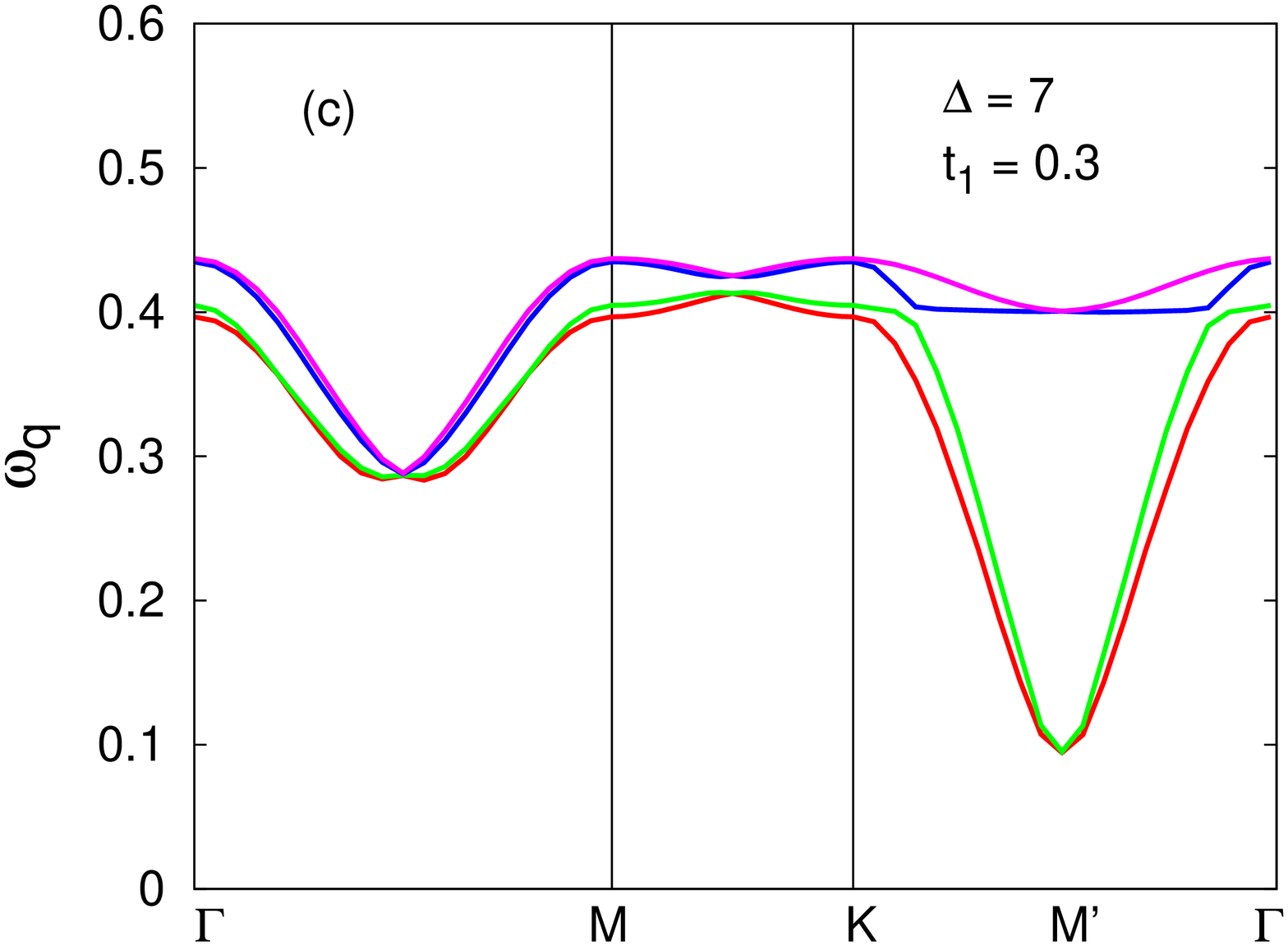,angle=-90,width=50mm} 
\caption{Stabilization of the zigzag order near ${\rm M'}$ due to finite-$U$-induced spin couplings (a,b) and emergence of finite gap even in the strong coupling limit with further suppression of the magnetic frustration (c).} 
\label{fig4}
\end{figure}

In order to suppress the instability near ${\rm M'}$, we will now study the effects of reduced hopping parameters ($t_n ^{(p)} < t_n$) for the frustrated (parallel) spins compared to the unfrusrated (antiparallel) spins, which results in net positive energy cost for the spin twisting mode (Appendix B). This reduced frustration results in only a marginal instability at ${\rm M'}$ as seen in Fig. \ref{fig4}(a) for hopping parameter values $t_1=0.4$, $t_1 ^{(p)}=0.3$, $t_2=0.25$, and $t_2 ^{(p)}=0.15$, in contrast to the strong instability seen in Fig. \ref{fig3}(b). 

Furthermore, this marginal instability in the strong coupling limit is removed at intermediate coupling, as seen in Fig. \ref{fig4}(b). This novel finite-$U$-induced stabilization of zigzag order is due to the second neighbor ferromagnetic spin couplings $J_2 ^{(p)}(U)$ induced at finite $U$ between parallel spins. This interaction also results in positive energy cost for the spin twisting mode from one domain to another (Appendix B). Further suppression of the frustration results in finite spin wave gap at ${\rm M'}$  even in the strong coupling limit, as shown in Fig. \ref{fig4}(c), where $t_1 = 0.3$, $t_1 ^{(p)}= 0.15$, $t_2=0.25$, $t_2 ^{(p)}=0.15$, and $\Delta=7$. 

\begin{figure}
\vspace*{0mm}
\hspace*{0mm}
\psfig{figure=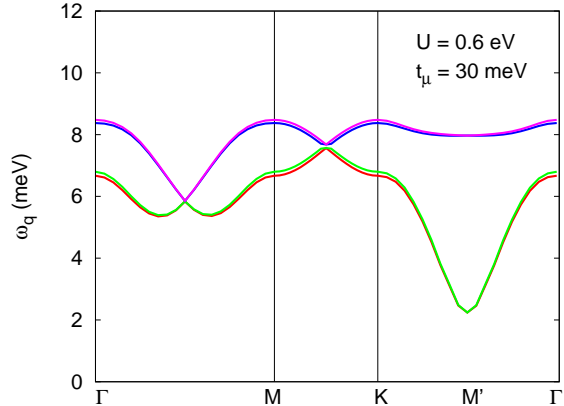,angle=-90,width=80mm} 
\caption{Calculated spin wave energies with realistic energy scales for the hopping and interaction terms. Here $t_1 = 0.1$, $t_2 = 0$, $\Delta = 10$ ($U\approx 20$), and energy scale $t_\mu=30$ meV.} 
\label{fig5}
\end{figure}

Finally, in order to investigate the magnetocrystalline anisotropy and easy axis, we will consider secondary spin-dependent hopping terms arising from structural distortions. We have therefore included small spin-dependent hopping terms $\tilde{t}_x$ ($\tilde{t}_y$) along the $y$ ($x$) bonds, so that $t_x \rightarrow (t_x, \tilde{t}_y)$ and $t_y \rightarrow (\tilde{t}_x, t_y)$. From the strong-coupling expansion (Appendix A), magnetocrystalline anisotropy and easy axis (along $S_z$ direction) is readily seen, along with the onset of symmetric off-diagonal spin interactions. Similarly, including secondary spin-dependent hopping terms along the $y,z$ ($z,x$) bonds will result in easy axis along the $S_x$ ($S_y$) direction in spin space. Figure \ref{fig5} shows the spin wave dispersion for $\tilde{t}_x = \tilde{t}_y = 0.2$, $t_1 = 0.1$, $t_2 = 0$ and $\Delta = 10$ ($U\approx 20$), with the energy scale $t_\mu=30$ meV. Features of the calculated spin wave dispersion are in qualitative agreement with the INS and RIXS measurements on $\rm \alpha-RuCl_3$ and $\rm Na_2 Ir O_3$.\cite{choi_PRL_2012,gretarsson_PRB_2013,ran_PRL_2017} 

For this choice of parameters, the value of $U \approx 0.6$ eV is similar to that obtained in spin wave studies of the one-band Hubbard model for $\rm Sr_2 Ir O_4$.\cite{iridate_paper} Using the transformation from the three-orbital basis to the $J_{\rm eff}$ basis, the intra-orbital Coulomb interaction $U_\mu = 3U \approx 2$ eV is in agreement with estimates from DFT studies. 

In the parameter regime where zigzag order is stable (positive spin wave energies over the entire BZ), we find that the stripy and N\'{e}el orders show negative spin wave energies over substantial parts of the BZ. This confirms the relative stability of the zigzag order due to the anisotropic magnetic interactions generated by the spin-dependent hopping terms arising from orbital mixing.

The essential features of the calculated spin wave dispersion are summarized here: (i) spin wave gap at $\Gamma$ due to the anisotropic Kitaev interactions, (ii) sinusoidally decreasing spin wave energy along the $\Gamma$ - M direction ($q_x=0,q_y=$ finite) with minimum near the midpoint, (iii) Goldstone mode at ${\rm M'}$ corresponding to the spin twisting from one zigzag domain to another,  and (iv) spin wave gap at ${\rm M'}$ and strong stabilization of zigzag order by secondary spin-dependent hopping terms which are responsible for magnetocrystalline anisotropy, easy axis. These essential features are in good agreement with inelastic neutron scattering measurements on $\rm \alpha - RuCl_3$ and $\rm Na_2 Ir O_3$. 

The honeycomb materials of interest adopt monoclinic $C2/m$ structure, and the structural distortion indicates presence of trigonal compression of the oxygen octahedra along the [1,1,1] direction. These distortions result in metal-ligand-metal bond angle different from 90$^\circ$, which can be as large as 100$^\circ$ in $\rm Na_2IrO_3$.\cite{ye_PRB_2012} The distortion-induced reduction in the local octahedral crystal field symmetry modifies the O-assisted hoppings which can introduce
small $t_\mu$ and $\tilde{t}_\mu$. Indeed, additional (secondary) orbital mixing terms arising from the trigonal distortion have been considered to account for the symmetric off-diagonal spin interactions.\cite{rau_PRL_2014,winter_JPCM_2017} In view of the substantial magneto-crystalline anisotropy, spin wave gap, and stabilization of zigzag order, the sensitive dependence on the secondary spin-dependent hopping terms highlights an interesting interplay between magnetic order and structural distortion. 

\subsection{Effective exchange interactions}

In order to stabilize the experimentally observed zigzag order in $\rm Na_2IrO_3$ and $\alpha-\rm{RuCl_3}$, the most appropriate spin model is by no means settled yet, including the sign of the nearest-neighbor Kitaev interaction. Within our one-band Hubbard model approach, the effective exchange interactions (Appendix A) are obtained as $J = 4(t^2 - t^2_\mu)/U$ and $K=8t^2_\mu/U$, which yield $J \approx -6$ meV and $K \approx 12$ meV for the parameter values considered in Fig. \ref{fig5}. These values of the exchange interactions with $J$ negative (FM) and $K$ positive (AFM) are in close agreement with previous spin wave studies for $\rm Na_2IrO_3$ and $\alpha-\rm{RuCl_3}$.\cite{chaloupka_PRL_2013,banerjee_NMAT_2016} 

On the other hand, within the strong coupling expansion carried out in the $t_{2g}$ manifold including Hund's coupling $J_H$ (assuming $U \gg J_H \gg \lambda \gg t$), the effective spin model in the $J_{\rm eff}=1/2$ sector yields $J$ positive (AFM) and $K$ negative (FM).\cite{rau_PRL_2014} Following the expressions derived for the exchange interactions, and substituting realistic values of the hopping terms in the three-orbital basis\cite{rau_PRL_2014} ($t_1 = 33$ meV, $t_2 = 270$ meV, and $t_3 = 27$ meV) and the interaction terms ($U = 1.7$ eV, $J_{\rm H} = 0.3$ eV) for $\rm Na_2IrO_3$,\cite{foyevtsova_PRB_2013, winter_PRB_2016} we obtain $J \approx 4$ meV, $K \approx -52$ meV and $\Gamma \approx 1$ meV, clearly showing $K$ as the dominant interaction term.
  

In the phase diagram plot in terms of the angles $\theta$ and $\phi$ [Fig. 2(a) in Rau {\it et al.}],\cite{rau_PRL_2014} where $\tan \theta = \sqrt{J^2 + K^2}/\Gamma$ and $\tan \phi = K/J$, one obtains $\theta \sim \pi/2$ (neglecting the SOD term $\Gamma$) and $\phi \ge 3 \pi/2$ (since $\tan \phi$ is large and negative), for which the stripy phase is shown as the ground state. Hence, no spin wave studies have been carried out with above realistic values of spin interactions, as the zigzag state is not the ground state without the $\Gamma$ term. For negative $K$, stable zigzag state is obtained only when a dominant $\Gamma$ term ($\Gamma > |K|$) is included.\cite{chaloupka_PRB_2015} With antiparallel spins on one bond and parallel spins on the other two, the zigzag state allows for energy lowering by the $\Gamma$ term, whereas the stripy phase (opposite situation) does not, accounting for the relative stability of the zigzag state even for negative $K$. However, $\Gamma > |K|$ is inconsistent with the realistic values of spin interactions as obtained above.


\section{Conclusions}

Spin waves in the zigzag-ordered antiferromagnetic state on a honeycomb lattice were investigated within a Hubbard model with spin-dependent hopping terms, focussing on the roles of the emergent Kitaev, Heisenberg, Dzyaloshinskii-Moriya, and symmetric-off-diagonal spin interactions on the stability of the zigzag order. While the anisotropic Kitaev interactions favor zigzag order, the DM interactions were found to destabilize this order by inducing spin canting. Highlighting the importance of structural distortions, the secondary spin-dependent hopping terms were shown to strongly stabilize the zigzag order. This is consistent with similar findings for symmetric-off-diagonal interactions in spin models and also account for magneto-crystalline anisotropy, easy axis, and spin wave gap. 

The calculated spin wave features are in good agreement with inelastic neutron scattering measurements on $\rm \alpha - RuCl_3$ and $\rm Na_2 Ir O_3$. The small magnitude of the spin-dependent hopping terms $t_\mu$ ($\sim 30$ meV) due to the avoided cancellation resulting from structural distortions accounts for the low spin wave energy scale measured in the honeycomb lattice compounds as compared to the appreciable magnitude of the NN orbital-mixing hopping term ($\sim 270$ meV) obtained from band structure studies. 


Our one-band model study focussing on the $J=1/2$ sector near the Fermi energy is justified by the electronic band structure for the minimal three-orbital model on the honeycomb lattice. With NN orbital mixing hopping terms and realistic value of spin-orbit coupling, the electronic band structure shows that the bands are grouped into well separated $J=1/2$ and 3/2 sectors, as indicated by the $J$-basis projection. Neglect of the mixing effects between the $J=1/2$ and 3/2 sectors is a
shortcoming of the one-band model. More generally, investigation of the interplay between the distinctive flat-band states of the minimal three-orbital model, the spin-orbit coupling which causes mixing between these states, and the zigzag-order staggered field which splits the $J=1/2$ sector bands should be of interest. 

\appendix
\section{Strong-coupling expansion and anisotropic spin interactions}

Starting with the Hubbard model (Eq. \ref{sp_dep_Hubbard}), the various anisotropic spin interactions (Kitaev, Dzyaloshinskii-Moriya, pseudo-dipolar, symmetric-off-diagonal) generated in the strong coupling expansion due to both the spin-dependent hopping terms $t_\mu$ and the normal hopping terms $t$ are obtained as:
\begin{eqnarray}
H_{\rm eff} ^{(2)} = & & \frac{{4t_x}^2}{U} \sum_{\langle ij \rangle}
\left [ S_i ^x S_j ^x - (S_i ^y S_j ^y + S_i ^z S_j ^z) - n_i n_j \right ] \\ \nonumber 
&+& \frac{{4t_y}^2}{U} \sum_{\langle ij \rangle} 
\left [ S_i ^y S_j ^y - (S_i ^z S_j ^z + S_i ^x S_j ^x) - n_i n_j \right ] \\ \nonumber 
&+& \frac{{4t_z}^2}{U} \sum_{\langle ij \rangle}  
\left [ S_i ^z S_j ^z - (S_i ^x S_j ^x + S_i ^y S_j ^y) - n_i n_j \right ] \\ \nonumber 
&+& \frac{8t{t_z}}{U} \sum_{\langle ij \rangle} ({\bf S_i} \times {\bf S_j}).\hat{z}
+ \frac{8t{t_x}}{U} \sum_{\langle ij \rangle} ({\bf S_i} \times {\bf S_j}).\hat{x}
+ \frac{8t{t_y}}{U} \sum_{\langle ij \rangle} ({\bf S_i} \times {\bf S_j}).\hat{y} \\ \nonumber
&+& \frac{8t_x t_y}{U} \sum_{\langle ij \rangle}  
\left ( S_i ^x S_j ^y + S_i ^y S_j ^x) \right ] 
+ \frac{8t_y t_z}{U} \sum_{\langle ij \rangle} 
\left ( S_i ^y S_j ^z + S_i ^z S_j ^y) \right ] 
+ \frac{8t_z t_x}{U} \sum_{\langle ij \rangle}
\left ( S_i ^z S_j ^x + S_i ^x S_j ^z) \right ] \\ \nonumber 
&+& \frac{4t^2}{U} \sum_{\langle ij \rangle} \left [ {\bf S_i} . {\bf S_j} -  n_i n_j \right ]
\end{eqnarray}

Here we have considered a general case where all three hopping terms $t_\mu$ are present for a given $\langle ij \rangle$ pair. The symmetric-off-diagonal terms $(S_i ^\alpha S_j ^\beta)$ are generated only if two components $t_\alpha$ and $t_\beta$ of the spin-dependent hopping terms are non-zero for the same pair $\langle ij \rangle$, and have been included in recent studies to stabilize the zigzag order.\cite{chaloupka_PRL_2013} 

\section{Zigzag domains and Goldstone mode}

\begin{figure}
\vspace*{0mm}
\hspace*{0mm}
\psfig{figure=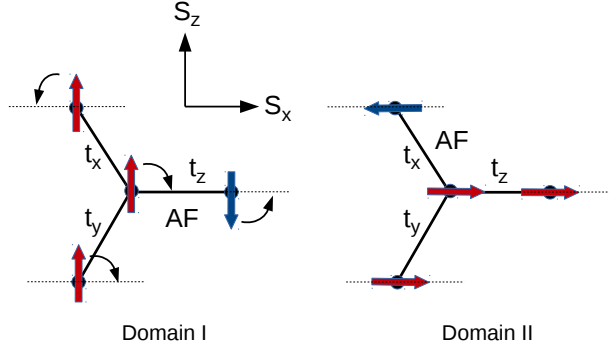,angle=0,width=80mm} 
\caption{Existence of 120$^\circ$-twinned magnetic domains for the zigzag order, connected by specific spin twisting modes as shown, results in the gapless Goldstone mode even in the presence of anisotropic spin interactions.} 
\label{domains}
\end{figure}

As seen from Fig. \ref{domains}, the AFM aligned spins correspond to the pair of sites with spin-dependent hopping $t_z$ ($t_x$) for the zigzag domain I (II). The specific spin twisting mode (Fig. \ref{domains}) allows for continuously transforming from one zigzag domain to another. The classical interaction energy is invariant under this continuous transformation as shown below, indicating the existence of a zero-energy Goldstone mode at the appropriate wave vector. 

For spin twisting by angle $\theta$ with respect to $S_z$ axis in the $S_z - S_x$ plane, the classical interaction energy for the pair of sites connected by hopping $t_z$ is obtained from the corresponding spin interaction terms:
\begin{equation}
E(t_z) = \frac{4{t_z}^2}{U} \left [ S_i ^z S_j ^z - (S_i ^x S_j ^x + S_i ^y S_j ^y) \right ] 
= \frac{4 {t_z}^2}{U} S^2 \left [ - \cos ^2 \theta - \sin ^2 \theta \right ] 
\end{equation}
which is independent of the twist angle $\theta$. The energy increase in the $z$ term is exactly compensated by the energy decrease in the $x$ term. Similar energy compensation results for the pair of sites connected by hopping terms $t_x$ and $t_y$. 

In contrast, for the isotropic exchange interactions $J_n = (4t_n ^2/U) {\bf S_i}.{\bf S_j}$, the energy invariance follows from an exact cancellation of the {\em different} contributions associated with a given site. If the exchange interactions between parallel and antiparallel spins are denoted by $J_n ^{(p)}$ and $J_n$ for first and second neighbors ($n=1,2$), the net energy cost:
\begin{equation}
\Delta E = \left (J_1 - J_1 ^{(p)} \right ) [1-\cos 2\theta] 
+ \left (J_2 - J_2 ^{(p)} \right ) [1-\cos 2\theta]
\end{equation}
which identically vanishes in the frustrated case $J_n ^{(p)} = J_n$. However, for reduced frustration $J_n ^{(p)} < J_n$ (or equivalently $t_n ^{(p)} < t_n$ in terms of the hopping parameters), the net energy cost becomes positive. 

\section*{Acknowledgement}
Helpful discussions with Jeroen van den Brink and Rajyavardhan Ray, and sponsorship grant from the Alexander von Humboldt Foundation for a research stay at IFW Dresden (AS) are gratefully acknowledged.

\end{document}